\documentclass[aps,prb,twocolumn,showpacs,amsmath,amssymb]{revtex4}
\usepackage{graphicx}
\usepackage{bm}

\def\br{ \bm{r} }
\def\bk{ \bm{k} }
\def\sign{ \,\mathrm{sign}\, }

\begin{document}

\title{Effects of interface spin-orbit coupling on tunneling between normal metal and chiral $p$-wave superconductor}

\author{S. Wu and K. V. Samokhin}

\affiliation{Department of Physics, Brock University, St. Catharines, Ontario L2S 3A1, Canada}

\date{\today}

\begin{abstract}
We study the tunneling conductance of a clean normal metal/chiral $p$-wave superconductor junction using the extended
Blonder-Tinkham-Klapwijk formalism. It is shown that the spin-orbit coupling of the Rashba type that is present near the interface causes the subgap conductance peaks associated with the Andreev surface bound states to shift to a nonzero bias.
We also investigate the effect of the Fermi wavevector mismatch between the normal metal and the superconductor.
\end{abstract}


\maketitle

\section{Introduction}

Tunneling spectroscopy is one of the most powerful probes of the electronic states in superconductors.
Quasiparticles in anisotropically-paired superconductors can experience a variation, e.g. a sign change, of the order parameter upon reflection from an interface. Then the interference between the incident and reflected quasiparticles results in the formation of bound states near the interface, with the energies inside the bulk energy gap, which are known as the Andreev bound states (ABS).\cite{Hu94,KashTan95} The ABS manifest themselves in low-energy features, typically peaks, in the tunneling conductance, which have been observed experimentally. In particular, the zero-bias conductance peaks in high-$T_c$ cuprates exhibit strong dependence on the crystallographic orientation of the interface, consistent with the $d$-wave pairing, see Ref. \onlinecite{ABS-review} for a review. The observation of broad subgap peaks in the tunneling conductance of Sr$_2$RuO$_4$ (Ref. \onlinecite{Mao01}) can also be explained in terms of the surface ABS, which are expected to exist in a chiral $p$-wave superconductor.\cite{YTK97,HonSig98} Other systems studied recently include the interfaces between magnetic or nonmagnetic normal metals and noncentrosymmetric or magnetic superconductors.\cite{LS07,Inio07,WuSam09} 
We would like also to mention that the zero-bias conductance peaks can also originate from the quasiparticle states localized near strong impurities or surface inhomogeneities.\cite{IBS-review} Those can be distinguished from the ABS by their different response on a magnetic field.\cite{SamWalk01}

Due to the breaking of reflection symmetry near the interface, quasiparticles experience the spin-orbit coupling (SOC) of the Rashba type,\cite{Rashba60} even if both the normal and superconducting crystals have inversion symmetry in the bulk. The effects of such interface SOC have been neglected in the previous studies of the tunneling conductance. In this paper we focus on the properties of a junction between a normal metal and a chiral $p$-wave superconductor. It is known that the ABS in this case correspond to Majorana fermions with linear dispersion, see Refs. \onlinecite{HonSig98,RG00}, and \onlinecite{SR04}. There are strong experimental indications that the chiral $p$-wave state is realized in Sr$_2$RuO$_4$ (Ref. \onlinecite{MM03}).
We neglect disorder and calculate the tunneling conductance using the Blonder-Tinkham-Klapwijk (BTK) formalism.\cite{BTK82} The effect of the Rashba SOC on the conductance can be attributed to a modification of the boundary conditions for the wavefunctions at the interface. We also take into account the difference between the Fermi wavevectors on the normal and superconducting sides.

The paper is organized as follows: In Sec. II, we develop a theoretical model of the normal metal-superconductor (N-S) junction with the interface SOC and use the BTK approach to calculate the amplitudes for various quasiparticle scattering processes. In Sec. III, the effects of both the interface SOC and the Fermi wavevector mismatch (FWM) on the tunneling conductance are presented and discussed. Sec. IV contains a summary of our
results. In Appendix, we analyze the ABS spectrum in a half-infinite chiral $p$-wave superconductor with an arbitrary interface potential. Throughout the paper we use the units in which $\hbar=1$.

\section{Formulation of the model}

\begin{figure}
\includegraphics[width=10cm]{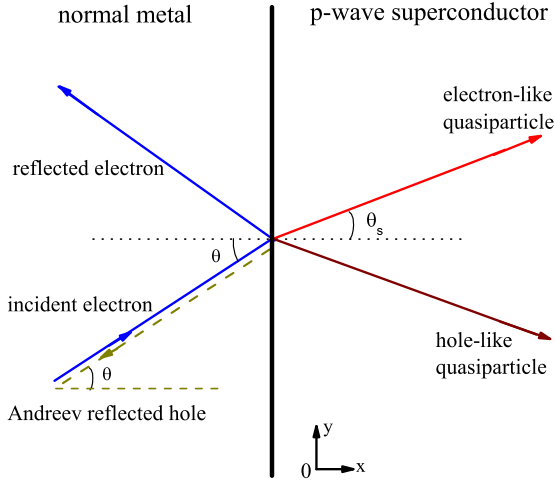}
\caption{(Color online) Schematic illustration of the quasiparticle reflection and
transmission processes at the N-S interface.} \label{fig: model}
\end{figure}

We consider a two-dimensional clean N-S junction shown in Fig. \ref{fig: model}. The interface is located at $x=0$ and is characterized
microscopically by a potential barrier which is not necessarily an even function of $x$, due to the crystal structure difference between the normal and superconducting sides. The asymmetric part of the potential is responsible for the interface SOC of the Rashba type. We consider the following model for the interface barrier:
\begin{equation}
\label{interface-model}
    U(x)=[U_{\textrm{0}}+U_{1}\bm{n}\cdot(\hat{\bm{\sigma}}\times\hat{\bk})]\delta(x),
\end{equation}
where $\bm{n}\equiv\hat{\bm{x}}$ is the unit vector along the interface normal, $U_0$ and $U_1$ are the strengths of the the spin-independent and the Rashba SOC contributions, respectively, $\hat{\bm{\sigma}}$ are the Pauli matrices, and $\hat{\bk}=-i\bm{\nabla}$. The band dispersions
are assumed to be parabolic, with the same effective masses of quasiparticles on both sides (according to Ref.
\onlinecite{YTI06}, the effect of the mass difference is equivalent to that caused by a variation of the interface potential
strength). On the superconducting side of the junction, we assume a chiral $p$-wave pairing state of the form
$\bm{d}(\bk)\propto \bm{\hat{z}}(k_{x}+ik_{y})$.

The Bogoliubov-de Gennes (BdG) equations for the four components of the quasiparticle wavefunction are decoupled into two independent pairs of two-component equations as follows:
\begin{equation}
\label{BdG-eqs-2}
    {\cal H}_\sigma\Psi(\br)=E\Psi(\br),
\end{equation}
where $\sigma=\pm$ for different spin orientations,
\begin{equation}
\label{H-BdG-sigma}
    {\cal H}_\sigma=\left(\begin{array}{cc}
    \displaystyle -\frac{\bm{\nabla}^{2}}{2m}-E_{Fi}+U_{\sigma}(x) & \Delta(\hat{\bk},\br)\\
    \Delta^{\dag}(\hat{\bk},\br) & \displaystyle \frac{\bm{\nabla}^{2}}{2m}+E_{Fi}-U_{\sigma}(x)
    \end{array}\right),
\end{equation}
and $U_{\sigma}(x)=(U_{0}-\sigma U_{1}\hat{k}_{y})\delta (x)$. The Fermi energies in the normal and
superconducting regions can be different due to different carrier densities, with $E_{Fi}=E_{FN}$ or $E_{FS}$.
The ratio of the corresponding wavevectors is characterized the dimensionless FWM parameter as follows: $\lambda_{0}=k_{FS}/k_{FN}=\sqrt{E_{FS}/E_{FM}}$.

The off-diagonal elements of Eq. (\ref{H-BdG-sigma}) contain the gap function $\Delta$. In the spirit of the BTK approach, we neglect self-consistency and assume the gap magnitude to be equal to $\Delta_0$ on the superconducting side, and zero on the normal side. Then we have
$\Delta(\hat{\bk},\br)=(\Delta_{0}/2k_{FS})\{(\hat k_{x}+i\hat k_{y}),\theta(x)\}$, where $\theta(x)$ is the step function.
The anticommutator on the right-hand side is required since the order parameter varies in space, see Ref. \onlinecite{SR04}. Below we use a simpler expression: $\hat\Delta=(\Delta_{0}/k_{FS})(\hat k_{x}+i\hat k_{y})$, at $x>0$, neglecting the $\delta$-function term in the off-diagonal elements, which gives a small correction to the boundary conditions.

Suppose an electron is injected from the normal metal with the excitation energy $E\geq 0$ and spin $\sigma$, at an angle $\theta$ from the interface
normal. The momentum parallel to the interface is conserved:
\begin{equation}
	k_{FN}\sin\theta=k_{FS}\sin\theta_s.
\end{equation}
The incident electron is reflected back either as an electron (normal reflection) or as a hole (Andreev reflection).\cite{And64}
In the superconductor, the wavefunctions of the transmitted quasiparticles have both electron and hole components.

Solution of Eq. (\ref{BdG-eqs-2}) has the form $\Psi(\br)=e^{ik_yy}\Psi(x)$, where
\begin{eqnarray}
\label{Psi_N}
    \Psi_N(x)&=&\left(\begin{array}{c}
    1\\ 0
    \end{array}\right)e^{ik_{FN}\cos\theta\, x}
    +a_{\sigma}
    \left(\begin{array}{c}
    0\\ 1
    \end{array}\right)e^{ik_{FN}\cos\theta\, x}
    \nonumber\\
    &&+b_{\sigma}
    \left(\begin{array}{c}
    1\\ 0
    \end{array}\right)e^{-ik_{FN}\cos\theta\, x}
    \end{eqnarray}
on the normal side, and
\begin{eqnarray}
\label{Psi_S}
    \Psi_{S}(x)&=&c_{\sigma}
    \left(\begin{array}{c}
    u\\ ve^{-i\theta_{s}}
    \end{array}\right)e^{ik_{FS}\cos\theta_{s}\,x}\nonumber\\
    &&+d_{\sigma} \left(\begin{array}{c}
    -v e^{-i\theta_{s}}\\ u
    \end{array}\right)e^{-ik_{FS}\cos\theta_{s}\,x}
\end{eqnarray}
on the superconducting side. Here $a_{\sigma}$ and $b_{\sigma}$ are the amplitudes of the Andreev and normal reflection, respectively,
and $c_{\sigma}$ and $d_{\sigma}$ are the transmission amplitudes. The electron and hole components of the wavefunctions in the
superconducting region are given by
\begin{eqnarray}
    u=\frac{1}{\sqrt{2}}\sqrt{1+\frac{\Omega}{E}},\qquad v=\frac{1}{\sqrt{2}}\sqrt{1-\frac{\Omega}{E}},
\end{eqnarray}
where $\Omega=\sqrt{E^{2}-\Delta_{\textrm{0}}^{2}}$.

All the reflection and transmission amplitudes in Eqs. (\ref{Psi_N}) and (\ref{Psi_S}) can be found from the 
boundary conditions that follow from Eq. (\ref{interface-model}):
\begin{equation}
\label{bc-1}
	\left.\begin{array}{l}
   	\Psi_S(0^{+})=\Psi_N(0^{-}),\\ \\
    	\Psi'_S(0^{+})-\Psi'_N(0^{-})\\
	\quad=2m(U_0-\sigma U_{1}k_{FN}\sin\theta)\Psi_N(0^{-}).\end{array}\right.
\end{equation}
In particular, for the reflection amplitudes we obtain:
\begin{equation}
    \left.\begin{array}{l}
    \displaystyle a_{\sigma}(E,\theta)=\frac{4\lambda
    \omega_{0}e^{-i\theta_{s}}}{(1+\lambda^{2}+Z_{\sigma}^{2})\omega_{+}+2\lambda\omega_{-}},\\ \\
    \displaystyle b_{\sigma}(E,\theta)=\frac{[(1-iZ_{\sigma})^{2}-\lambda^{2}]
    \omega_{+}}{(1+\lambda^{2}+Z_{\sigma}^{2})\omega_{+}+2\lambda\omega_{-}},
    \end{array}\right.
\label{eq:coeffiecents}
\end{equation}
where
\begin{eqnarray*}
    &&\lambda=\lambda_{0}\frac{\cos\theta_{s}}{\cos\theta},\\
    &&\omega_{0}=\frac{u}{v},\quad \omega_{+}=\omega_{0}^{2}+e^{-2i\theta_{s}},\quad
    \omega_{-}=\omega_{0}^{2}-e^{-2i\theta_{s}},\\
    &&Z_{\sigma}=\frac{Z_{0}}{\cos\theta}-\sigma Z_{1}\frac{\sin\theta}{\cos\theta},\quad
    Z_0=\frac{2mU_0}{k_{FN}},\quad Z_1=2mU_1.\quad
\end{eqnarray*}
The dimensionless parameters $Z_0$ and $Z_1$ characterize the strengths of the potential and SO scattering, respectively.

From the conservation of the probability current it follows that
$$
 	|a_\sigma|^2+|b_\sigma|^2+\lambda(|c_\sigma|^2+|d_\sigma|^2)=1.
$$
We note that the charge transmission at subgap energies is enhanced when the normal reflection is minimized, i.e. at $b_\sigma=0$, which happens if $\omega_{+}=0$.
Then Eq. (\ref{eq:coeffiecents}) yields a subgap resonance with the energy $E=\Delta_{0}\sin\theta_s$, $\theta_s>0$, corresponding to a branch of excitations localized near the interface. This result agrees with that of Refs. \onlinecite{HonSig98,RG00}, and \onlinecite{SR04}, see also Appendix.

By using the BTK formalism,\cite{BTK82} we obtain for the dimensionless angle-resolved differential tunneling conductance:
\begin{widetext}
\begin{eqnarray}
    G_{\sigma}(E,\theta)&=&1+|a_{\sigma}(E,\theta)|^{2}-|b_{\sigma}(E,\theta)|^{2} \nonumber\\
    &=&\frac{4\lambda\{[(1+\lambda)^{2}+Z_{\sigma}^{2}]\omega_{0}^{4}+4\lambda\omega_{0}^{2}-[(1-\lambda)^{2}
    +Z_{\sigma}^{2}]\}}
    {[(1+\lambda)^{2}+Z_{\sigma}^{2}]^{2}\omega_{0}^{4}+2\cos 2\theta[(1+\lambda)^{2}+Z_{\sigma}^{2}][(1-\lambda)^{2}
    +Z_{\sigma}^{2}]\omega_{0}^{2}+[(1-\lambda)^{2}+Z_{\sigma}^{2}]^{2}}\label{eq:conductance}.
\end{eqnarray}
\end{widetext}
One can see that, while the conductance depends on the incident spin orientation: $G_+(E,\theta)\neq G_-(E,\theta)$, the time reversal invariance is respected: $G_\sigma(E,\theta)=G_{-\sigma}(E,-\theta)$.
The experimentally measurable tunneling conductance $G(E)$ is obtained after the angular integration and the summation over the incident spin orientations
as follows:
\begin{equation}
\label{G-integrated}
    G(E)=\frac{1}{G_{N}}\sum\limits_{\sigma}\int_{-\pi/2}^{\pi/2} d\theta\cos\theta\,
    G_{\sigma}(E,\theta).
\end{equation}
Here $G_{N}$ is the conductance for a normal metal/normal metal junction with the interface potential given by Eq. (\ref{interface-model}):
\begin{equation}
\label{GN-integrated}
    G_{N}=\sum\limits_{\sigma}\int_{-\pi/2}^{\pi/2} d\theta\cos\theta\,
    G_{N\sigma}(E,\theta),
\end{equation}
with
\begin{equation}
\label{G_N-st}
	G_{N\sigma}(E,\theta)=\frac{4}{4+Z_{\sigma}^{2}}.
\end{equation}
We note that the difference between the Fermi wavevectors imposes a constraint on the effective range of angles contributing to the integrals in Eqs. (\ref{G-integrated}) and (\ref{GN-integrated}): if $k_{FS}<k_{FN}$, then there is no transmission for $|\sin\theta|>\lambda_{0}$.

\section{Results}

The tunneling conductance of the N-S junction at zero temperature can be plotted as a function of the dimensionless
excitation energy $E/\Delta_{\textrm{0}}$. We will study the effects of the interface SOC and the difference between the Fermi wavevectors on the tunneling spectra.

\begin{figure}
\includegraphics[width=7.5cm]{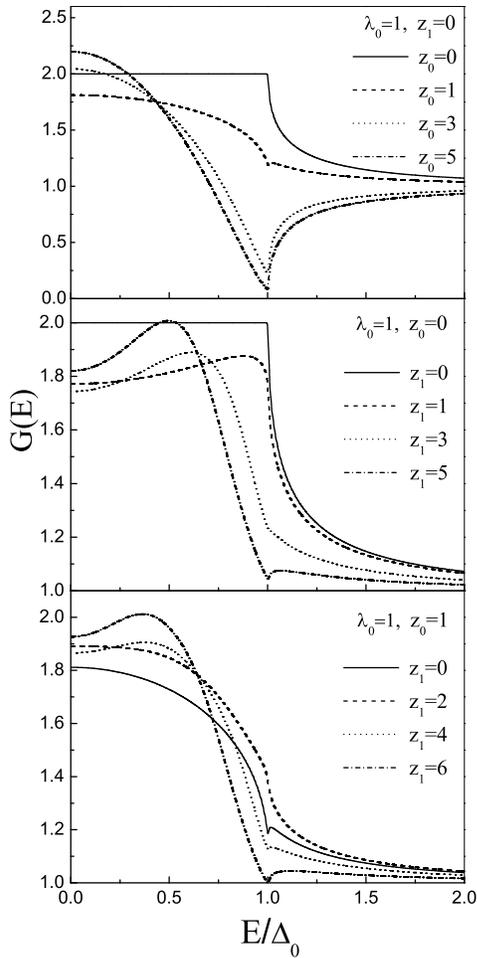}
\caption{The dimensionless tunneling conductance $G(E)$ versus
$E/\Delta_{0}$ for $\lambda_{0}=1$ (no FWM) and different values of $Z_{0}$
and $Z_{1}$: $Z_{1}=0$ (top panel), $Z_{0}=0$ (middle panel), and
$Z_{0}=1$ (bottom panel).}
\end{figure}

Let us first consider the case in which there is no FWM, i.e. $k_{FN}=k_{FS}$ and $\lambda_{0}=1$. Fig.~2
shows the tunneling conductance $G(E)$ for different relative strengths of the purely potential and the SOC contributions to the interface scattering.
In the top panel, we show $G(E)$ in the absence of the interface SOC ($Z_{1}=0$).
The middle and bottom panels demonstrate the effects of varying the interface SOC at $Z_{0}=0$ (high transparency barrier) and $Z_0=1$ (low transparency barrier). Note that, if $Z_{0}=Z_{1}=0$, then $G(E)=2$ for all subgap energies, $E<\Delta_{0}$, due to the Andreev reflection occuring with the probability one.

The broad peak at the subgap energies, which is most pronounced for a low-transparency interface, is associated with the surface ABS that exist in chiral $p$-wave superconductors\cite{HonSig98} (we recall that there are no subgap surface bound states in $s$-wave superconductors, and the tunneling conductance at $E<|\Delta|$ is strongly suppressed for large $Z_0$, Ref. \onlinecite{BTK82}). When we include the interface SOC, the conductance peak is shifted to a nonzero bias, see the bottom panel of Fig.~2. The origin of this effect can be attributed to a nonmonotonic angular dependence of the transmission coefficients at $Z_1\neq 0$, which is evident from Eq. (\ref{G_N-st}). Note that the spectrum of the surface ABS is not affected by the interface SOC, see Appendix.

\begin{figure}
\includegraphics[width=7cm]{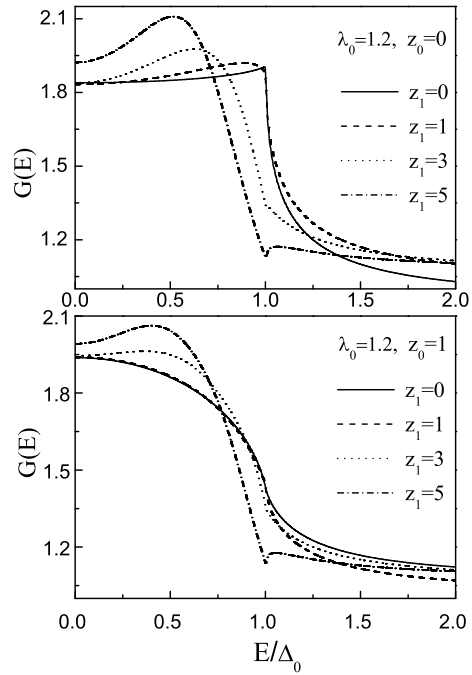}
\caption{The dimensionless tunneling conductance $G(E)$ versus
$E/\Delta_{0}$ for the FWM parameter $\lambda_{0}=1.2$ and different values of the interface SOC.
In the top panel $Z_{0}=0$ (no potential barrier), in the bottom panel $Z_{0}=1$ (strong potential barrier).}
\end{figure}

\begin{figure}
\includegraphics[width=7cm]{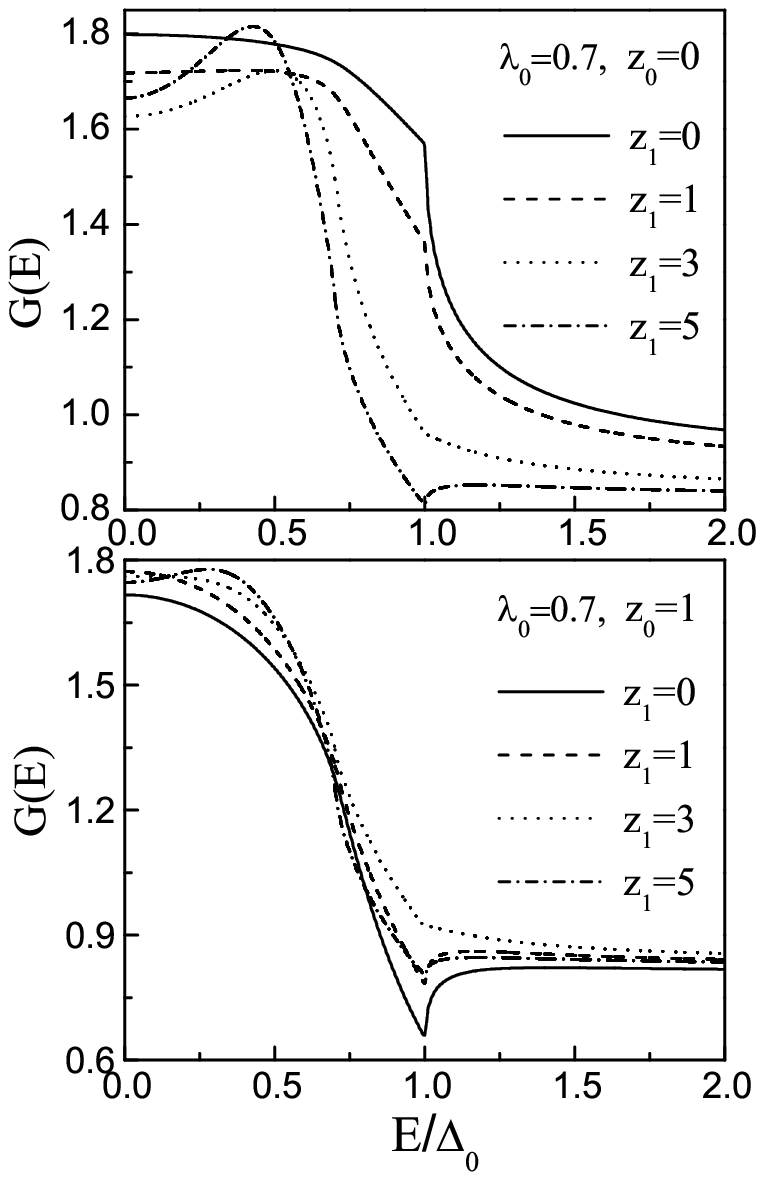}
\caption{The dimensionless tunneling conductance $G(E)$ versus
$E/\Delta_{0}$ for the FWM parameter $\lambda_0=0.7$. Other parameter values are the
same as in Fig.~3.}
\end{figure}

Next, we discuss the effect of the FWM on the tunneling conductance, in two cases: (i) $\lambda_{0}=1.2$, (ii) $\lambda_{0}=0.7$.
In each case, we consider both high- and low-transparency interfaces, $Z_{0}=0$ and $Z_{0}=1$. Figs. 3 and 4 show the variation of $G(E)$ for several values of $Z_{1}$. In the case (ii) (and for $k_{FS}<k_{FN}$, in general), the conductance is notably suppressed by the FWM, because the modes with $|\sin\theta|>\lambda_{0}$ experience total reflection and, therefore, do not contribute to the conductance.
In contrast, there is no discernible consistent effect of the FWM in the case (i).

\section{Summary}

In summary, we have applied the extended BTK formalism to investigate the tunneling conductance of a junction between a normal metal and a chiral $p$-wave superconductor. We focused on the effects of the interface SOC of the Rashba type, both with and without the Fermi surface mismatch between
the two sides of the junction. The structure of the subgap peaks in the tunneling conductance has been shown to strongly depend on the interface SOC. In particular, 
for a low-transparency interface, the maximum of the conductance associated with
the surface ABS is shifted away from the zero bias. We also considered the case of different Fermi wavevectors in the normal and superconducting regions. When $k_{FS}<k_{FN}$, the tunneling conductance is strongly suppressed.

\section*{Acknowledgements}

This work was supported by a Discovery Grant from the Natural
Sciences and Engineering Research Council of Canada.

\appendix
\section{Surface ABS with spin-orbit coupling}

Let us consider a superconductor with a flat surface, occupying the $x\geq 0$ half-space. We assume a hard-wall confining potential $V(x)$, which is infinite at $x<0$ and varies within a thin surface layer, whose thickness $a$ is of the order of several lattice spacings. This potential results in the SOC of the form $(1/4m^2c^2) V'(x)(\hat{\bm{p}}\times\hat{\bm{\sigma}})_x$, which is also restricted to the vicinity of the surface (recall that $\hbar=1$ in our units). Assuming a 2D geometry with the band dispersion $\xi(\bk)$ (which includes the chemical potential), the quasiparticle spectrum can be found from the following $4\times 4$ BdG Hamiltonian:
\begin{equation}
\label{H-BdG-general}
	{\cal H}=\left(\begin{array}{cc}
 	\hat\epsilon & \hat\Delta\\
	\hat\Delta^\dagger & -\hat\epsilon^T
	\end{array}\right),
\end{equation}
where $\hat\epsilon=\xi(-i\bm{\nabla})+V(x)-(i/4m^2c^2)V'(x)\hat\sigma_3\nabla_y$ is the single-particle Hamiltonian. In a triplet pairing state, the order parameter is given by $\hat\Delta=(i\hat{\bm{\sigma}}\hat\sigma_2)\bm{d}(\hat{\bk},\br)$. To make analytical progress, we neglect self-consistency and assume that the order parameter is uniform at all $x>a$. The system is translationally invariant along the surface and the wavefunctions have the form $\Psi(\br)=e^{ik_yy}\Psi(x)$.

For a chiral $p$-wave state of the form $\bm{d}\propto\bm{\hat{z}}(k_{x}+ik_{y})$, it is easy to show that Eq. (\ref{H-BdG-general}) can be written as a direct sum of two $2\times 2$ Hamiltonians given by
\begin{equation}
\label{H-pm}
	{\cal H}_\sigma=\left(\begin{array}{cc}
 	\hat\xi+U_\sigma(x) & \hat\Delta\\
	\hat\Delta^\dagger & -\hat\xi-U_\sigma(x)
	\end{array}\right),
\end{equation}
$\sigma=\pm$, $U_\sigma(x)=V(x)-\sigma(1/4m^2c^2)V'(x)k_y$, and the gap function is given by $\hat\Delta=(\Delta_{0}/k_F)(\hat k_{x}+ik_{y})$, with $k_F\equiv k_{FS}$ being the Fermi wavevector.

In the region $x>a$, $U_\sigma(x)=0$, and the spectra of ${\cal H}_\sigma$ can be analyzed in the semiclassical, or Andreev, approximation.\cite{And64} We represent wavefunctions in the form $\Psi(x)=e^{ik_xx}\psi(x)$, where $k_x$ satisfies the equation
\begin{equation}
\label{xi-eq}
	\xi(k_x,k_y)=0,
\end{equation}
at fixed $k_y$. Each Fermi-surface wavevector $\bk=(k_x,k_y)$ defines a semiclassical trajectory, along which the quasiparticle state is described by a coherent superposition of the electron and hole amplitudes: $\psi(x)=[\psi_e(x),\psi_h(x)]^T$, satisfying the Andreev equation:
\begin{equation}
\label{And-eq-gen}
	\left(\begin{array}{cc}
		-iv_{F,x}(\bk)\nabla_x & \Delta(\bk) \\
		\Delta^*(\bk) & iv_{F,x}(\bk)\nabla_x
	\end{array}\right)\psi(x)
	=E\psi(x),
\end{equation}
where $\bm{v}_F(\bk)=\partial\xi/\partial\bk$ is the quasiparticle velocity on the Fermi surface, $E\geq 0$ is the energy of excitations, and
\begin{equation}
\label{chiral-p-wave}
	\Delta(\bk)=\Delta_0\frac{k_x+ik_y}{k_F}.
\end{equation}
Depending on the direction of propagation, the semiclassical trajectories are classified as either incident ($v_{F,x}<0$) or reflected ($v_{F,x}>0$).

One can seek solution of Eq. (\ref{And-eq-gen}) in the form of a plane wave:
$\psi(x)\sim e^{iqx}$. Focusing on the quasiparticle states which are bound to the surface, but cannot exist in the bulk, we expect that $E\leq\Delta_0$. For the wavefunction we then obtain:
\begin{equation}
\label{Andreev psi}
	\psi_{\bk}(x)=\frac{1}{\sqrt{2}}
	\left(\begin{array}{c}
		1 \\ \alpha(\bk)
	\end{array}\right)
	e^{-\kappa(\bk)x},
\end{equation}
where
$$
	\alpha(\bk)=\frac{\Delta^*(\bk)}{E+i\Omega(\bk)\sign v_{F,x}(\bk)},\quad \kappa(\bk)=\frac{\Omega(\bk)}{|v_{F,x}(\bk)|},
$$
and $\Omega(\bk)=\sqrt{|\Delta(\bk)|^2-E^2}=\sqrt{\Delta_0^2-E^2}$.	

The Andreev approximation is not valid near the surface, at $0<x<a$, where the rapidly varying potential $U_\sigma(x)$ is nonzero. The surface scattering will result in the effective boundary conditions at $x=a$, which express the Andreev wavefunctions corresponding to the reflected trajectories in terms of those corresponding to the incident trajectories.\cite{Shel-bc}

Depending on the band structure and the surface orientation, Eq. (\ref{xi-eq}) might have several solutions. In the case of a parabolic band, there is only one incident and one reflected trajectory at each $k_y$ (except $k_y=\pm k_{FS}$, where the Andreev approximation is not applicable): $\bk_{in}=(-k_x,k_y)$, $\bk_{out}=(k_x,k_y)$, with $k_x>0$. The effective boundary condition has the following form:
\begin{equation}
	\psi_{\bk_{out}}(a)=\hat S\psi_{\bk_{in}}(a),
\end{equation}
where $\hat S$ is the surface scattering matrix, which is an electron-hole scalar\cite{Shel-bc} determined by the surface potential $U_\sigma(x)$. Inserting here the wavefunctions (\ref{Andreev psi}), we arrive at an equation for the bound state energy:
\begin{equation}
\label{E-eq}
	\frac{E+i\Omega(\bk_{in})}{E-i\Omega(\bk_{out})}=\frac{\Delta(\bk_{in})}{\Delta(\bk_{out})}.
\end{equation}
For the chiral $p$-wave state (\ref{chiral-p-wave}), the solution is given by
\begin{equation}
\label{Majorana-spectrum}
	E(k_y)=\Delta_0\frac{k_y}{k_F} \quad (k_y>0).	
\end{equation}
Remarkably, this expression does not contain any microscopic details (in particular, it has the same form for ${\cal H}_+$ and ${\cal H}_-$), which is consistent with a topological nature of the surface ABS.\cite{Volovik97}
We come to the conclusion that the surface bound states of the Hamiltonian (\ref{H-BdG-general}) are described by two ($\sigma=\pm$) degenerate branches of fermionic excitations with linear dispersion, given by Eq. (\ref{Majorana-spectrum}), regardless of the SOC strength.

\end{document}